# Dielectric Properties of Polysulfone – Carbon Nanotube Composite Membranes

Bhakti Hirani
*Department of Biotechnology*
*Pillai College of Arts, Commerce & Science (Autonomous)*
New Panvel, India
bhaktih@mes.ac.in

P. S. Goyal
*Department of Electronics*
*Pillai College of Engineering*
New Panvel, India
psgoyal@mes.ac.in

Deepali Shrivastava
*Department of Applied Chemistry*
*Pillai College of Engineering*
New Panvel, India
deepali@mes.ac.in

S. K. Despande
*UGC-DAE Consortium for Scientific Research*
*Bhabha Atomic Research Centre (BARC)*
Mumbai, India
skdesh@csr.res.in

*Abstract* — **Polymeric membranes, including Polysulfone (PSf) membranes, are routinely used for water treatment. To enhance water permeation of above membranes, it is common to synthesize polymeric membranes with carbon nanotubes (CNTs) embedded in them. It is seen that water permeability of membranes having vertically aligned CNTs is higher, as compared to those where CNTs are not aligned. It is of interest to examine if the dielectric constant of a CNT based nanocomposite membrane is sensitive to alignment of CNTs or not. This paper reports dielectric properties of PSf-MWCNT membranes, both, for aligned and unaligned MWCNTs. Multi Walled Carbon Nanotubes (MWCNTs) based polysulfone membranes were synthesized using standard methods. MWCNTs in above membranes were aligned by casting the membrane in presence of magnetic field. Dielectric parameters ($\varepsilon'$, $\varepsilon''$, and $\delta$) for above membranes were measured over a frequency range of 10Hz to 1MHz. It was seen that there is significant change in real part $\varepsilon'$ of dielectric constant depending on whether membrane is casted without or with magnetic field. The changes in imaginary part $\varepsilon''$ of dielectric constant and the dielectric loss factor tan $\delta$ are small. It may be mentioned that it is known in literature that $\varepsilon'$ increases when CNTs are aligned in conventional CNT – polymer nanocomposites. The present paper, for the first time, shows that the above result is valid for membranes also.**

I. INTRODUCTION

Dielectric materials are widely used in electrical and electronic industry [1,2]. Depending on the application, one may need low dielectric materials or high dielectric materials. For example, materials having low dielectric constant are used for packaging electronic systems [3], and materials having high dielectric constant are used for energy storage devices [4]. In both the cases, it is desirable to use polymeric materials as they have several attractive properties such as low density, reasonable strength, flexibility, easy processability, etc. More importantly, dielectric constant of a polymer can be easily altered by adding suitable additives. It is common to enhance dielectric constant of polymeric materials by blending them with ceramics such as $BaTiO_3$ or conductive fillers such as carbon nanotubes (CNTs) [4 - 6]. The subject of CNT-Polymer nanocomposites is of wide interest and a recent book [6] gives an excellent discussion on their properties. including electrical properties, of these materials. It is seen that electric and dielectric properties of CNT based composites are different depending on whether CNTs are aligned or not. This is true both for single walled carbon nanotubes (SWCNTs) and multiwall carbon nanotubes (MWCNTs).

It is of interest to know if dielectric constant of CNT based polymeric membranes depends on alignment of CNTs. It may be mentioned that membranes are different from conventional polymer films in the sense that polymeric membranes have porous structure. Though there are many studies which deal with dielectric properties of Polymer – CNT nanocomposite [7, 8]; such studies on polymeric membranes are limited. The present paper deals with dielectric properties of Polysulfone-MWCNTs composite membranes, both, for aligned and un-aligned MWCNTs.

The membranes are porous polymeric films and are used for water treatment etc. [9, 10]. There is considerable interest in incorporating CNTs in above membranes as presence of CNTs improves their water permeability [11, 12]. Our group have carried out series of experiments on MWCNT based polysulfone membranes to understand their water permeation properties [13-15]. It was seen that water permeability of MWCNT based polysulfone membranes improves if one aligns the MWCNTs in a direction perpendicular to the surface of the film. This paper deals with the dielectric properties of above membranes. In particular, it examines if the dielectric constant of MWCNT based membranes increases on alignment of MWCNTs.

## II. Experimental

### A. Synthesis of Pure Poplysulphone Membranes

Pure polysulfone membrane was synthesized using polysulfone (PSf) as base polymer, polyvinyl pyrrolidone (PVP) as pore former, N methyl - 2 - pyrrolidone (NMP) as solvent using a method known as "phase inversion" method [16]. 25 gm of PSf beads and 7.5 gm of PVP were dissolved in 87.5 ml NMP and the solution was agitated vigorously for several hours to attain complete dissolution of PSf. The dope solution so obtained was kept at room temperature (about 27°C) for two days to eliminate the trapped bubbles from the mixture. The membrane was synthesized from dope solution using the procedure illustrated in Fig.1. The dope solution was poured on casting plate (1) and this plate was allowed to pass underneath a casting blade (2) and then between the two magnetic poles (3, 4) and finally it was submerged in water tank (5). The film of viscous polymer solution precipitates in water to form a thin polymer sheet, and this sheet is referred to as membrane. In practice, the above steps were performed using an indigenously developed membrane casting machine [17]. This machine allows casting of membrane in presence of magnetic field, and also has a provision to choose membrane thickness in range of 50 to 500 µm. Magnet was kept off while synthesizing pure PSf membrane. The thickness of the membrane was kept at 100 µm.

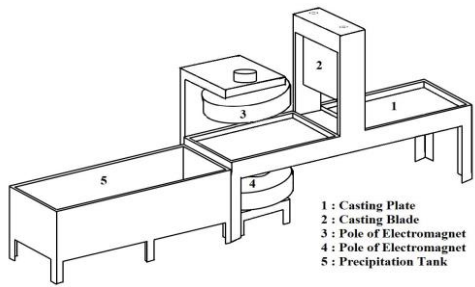

Fig 1. Schematic representation of membrane casting set-up

### B. Synthesis of PSf - MWCNTs Composite Membranes having aligned and un-aligned MWCNTs

MWCNT based PSf membranes were casted using same procedure as used for pure PSf membranes; the only difference being that dope contained MWCNTs. In general, nanotubes in above membranes would have random orientations. However, it is possible to align MWCNTs if, in addition to MWCNTs, the polymer solution contains nanomagnetic particles and the membranes are casted in presence of magnetic field [15, 17]. The magnetic field aligns magnetic particles along the direction of the magnetic field, and their alignment in turn, aligns nano-tubes.

In view of above 0.1 gm of MWCNTs and 1.0 gm of nanoparticles of $Fe_3O_4$ were added to dope while synthesizing composite membranes. MWCNTs (outer diameter ~ 6 – 13 nm, and length ~ 2.5 – 20 µm) were procured from M/s. United Nanotech Innovations Pvt. Ltd. India and nanoparticles (size ~ 22 nm) of $Fe_3O_4$ were synthesized in our laboratory [18]. The membranes were casted using above mentioned membrane-casting machine where there is a provision to cast the membrane in absence or presence of magnetic field. Magnetic field of 1500 Gauss was used. Depending on whether a membrane is casted in presence or absence of magnetic field, membranes having aligned or unaligned MWCNTs were synthesized. The membranes having un-aligned nanotubes are referred to as "PSf + MWCNTs - Without Field" and membranes having aligned nanotubes are referred to as "PSf + MWCNTs - With Field". It may be mentioned that nanotubes in above membranes were aligned in a direction perpendicular to film surface.

### C. Characterization of Polysulphone - MWCNTs Composite Membranes

The membranes were characterized by measuring their water permeability. The permeabilities of pure PSf membrane, (PSf / MWCNTs - Without Field) membrane and (PSf / MWCNTS - With Field) membrane were measured using a cross-flow filtration cell [13, 15]. Measurements were made on a circular membrane (~ 35 mm diameter), corresponding to a water pressure difference of 2 bar across the membrane. Volume of permeated water was collected for 5 minutes and average of three readings were used to obtain the permeation flux in units of Liters/square meter/hour (referred to as $LM^{-2}H^{-1}$).

### D. Dielectric Experiments

The dielectric constant of polymeric materials is a complex number that describes the interaction of the material with oscillating electric field. That is,

$$\varepsilon = \varepsilon' - j\varepsilon'' = \varepsilon'(1 - j\tan\delta) = |\varepsilon| e^{-j\delta} \quad \ldots\ldots\ldots (1)$$

where $\varepsilon'' = \varepsilon' \tan\delta$

Here $\varepsilon'$ is the real part of dielectric constant, $\varepsilon''$ is imaginary part of the dielectric constant and $\tan\delta$ is the dielectric loss factor. The dielectric parameters ($\varepsilon'$, $\varepsilon''$, and $\delta$) for pure PSf membrane, (PSf / MWCNTs - Without Field) membrane and (PSf / MWCNTS - With Field) membrane were measured over a frequency range of 10Hz to 1MHz using and impedance analyzer (Novocontrol, Germany model Alpha-AN) with the membrane sample sandwiched between electrodes in a parallel plate capacitor configuration. Measurements were made at room temperature and sample thickness was 100 µm.

## III. RESULTS AND DISCUSSION

### A. Water Permeation Studies

The measured permeation flux for pure PSf membrane, (PSf / MWCNTs - Without Field) membrane and (PSf / MWCNTS - With Field) membrane were 96 LM$^{-2}$H$^{-1}$, 145 LM$^{-2}$H$^{-1}$ and 577 LM$^{-2}$H$^{-1}$ respectively. As discussed below, these results suggest that nanotubes in (Polysulfone/ MWCNTS - With Field) are aligned in direction of water flow.

The water permeability of pure PSf membrane is decided by the pores in the membrane. However, once MWCNTs are embedded in PSf films, they provide additional paths for water flow. This explains increase in permeation flux from 96 LM$^{-2}$H$^{-1}$ to 145 LM$^{-2}$H$^{-1}$ in going from pure PSf to (PSf / MWCNTs - Without Field. This is in spite of the fact that MWCNTs in above membrane are oriented randomly. A large increase in permeation flux from 145 LM$^{-2}$H$^{-1}$ to 577 LM$^{-2}$H$^{-1}$ in going from (PSf /MWCNTs -Without Field) membrane to (PSf / MWCNTs - With Field) membrane is connected with alignment of MWCNTs. That is, the magnetic field aligns the MWCNTs in direction of water flow and that gives rise to increase in water permeability. The above water permeation studies clearly show that nanotubes are randomly oriented in PSf /MWCNTs -Without Field) membrane and they are vertically oriented in (PSf / MWCNTs - With Field) membrane. The present data, however, do not provide information on degree of alignment of nanotubes. No effort was made to obtain above information using advanced techniques such as Transmission Electron Microscopy (TEM) or Small Angle X-ray Scattering (SAXS).

### B. Dielectric Studies

The measured variation of ε' and ε" with frequency for above three membranes are shown in Figures 2 and 3 respectively. The data at high frequencies are more reliable as compared to those at low frequencies (see next paragraph). In view of that data are plotted in order of decreasing frequency, i.e., starting from 1 MHz and ending at 10 Hz. Further it may be noted that to avoid big numbers, frequency axis is plotted on log scale. Tables I, II and III give values of ε' and ε" and tan δ respectively for the three membranes at selected frequencies. It is seen that change in dielectric parameters in going from pure PSf membrane and (PSf / MWCNTs - Without Field) membrane are small. It is not surprising as nanotubes are not aligned in above membranes. It is seen that changes are much more significant in going from membrane casted without field to the one casted in field. Fig.2 shows that there is about 20% increase in real part of dielectric constant in going from (PSf / MWCNTs – Without Field) membrane to (PSf / MWCNTs – With Field) membrane at high frequencies ($10^3 – 10^6$ Hz), the frequency region where data does not suffer from superius effects.

In all the plots, the sharp increase in permittivity with lowering frequency below about 100Hz is due to the electrode polarization effects, arising due to charge accumulation at the interface between the electrodes and the sample. This effect diminishes at frequencies >100Hz since the charges are unable to respond to these frequencies. The permittivity ε' shows a significant enhancement when the MWCNTs are aligned. In the absence of the magnetic field, the Fe$_3$O$_4$ particles are randomly dispersed and since MWCNTs are not aligned, the effective area of the MWCNT "electrodes" may be lower and the spacing random. There is also the likelihood of conducting pathways due to MWCNTs, but at small concentrations of MWCNT, this effect may be small. Upon application of the field, the effective area of the MWCNTs would increase due to alignment of the MWCNTs, and this leads to increased polarization (capacitance) and hence enhanced permittivity of the measured specimen. In short, above studies show that there is large increase in ε' when nanotubes get aligned. Interestingly, application of the magnetic field shows no appreciable effect on ε", especially at higher frequencies. For example, at 10kHz, the tan δ is about 0.025 for both cases - with field as well as without field. This means that the alignment of MWCNTs in these samples leads to enhanced permittivity with low loss.

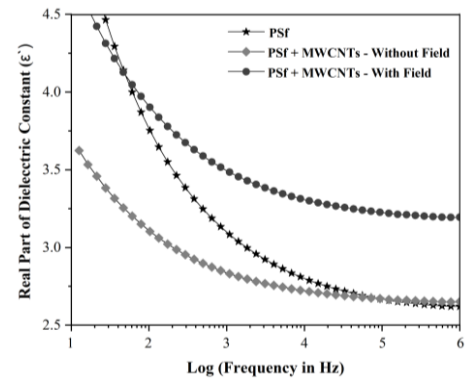

Fig 2. Variation of real part of dielectric constant (ε`) vs frequency

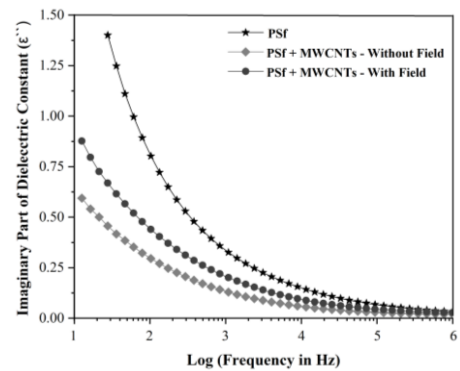

Fig 3. Variation of. imaginary part of dielectric constant (ε``) vs frequency

TABLE I   Real Part of Dielectric Constant (ε`) at Specific Frequencies

| Sr. No. | Real Part of Dielectric Constant (ε`) | | | |
|---|---|---|---|---|
| | *Frequency* | *PSf* | *PSf + MWCNTs – Without Field* | *PSf + MWCNTs – With Field* |
| 1. | **1000KHz** | 2.62 | 2.65 | 3.20 |
| 2. | **100KHz** | 2.65 | 2.67 | 3.22 |
| 3. | **10KHz** | 2.78 | 2.72 | 3.30 |
| 4. | **1KHz** | 3.08 | 2.83 | 3.48 |
| 5. | **100Hz** | 3.75 | 3.10 | 3.90 |
| 6. | **10Hz** | 5.33 | 3.70 | 4.79 |

TABLE II   IMAGINARY PART OF DIELECTRIC CONSTANT (ε``) AT SPECIFIC FREQUENCIES

| Sr. No. | Imaginary Part of Dielectric Constant (ε``) | | | |
|---|---|---|---|---|
| | *Frequency* | *PSf* | *PSf + MWCNTs – Without Field* | *PSf + MWCNTs – With Field* |
| 1. | **1000KHz** | 0.0327 | 0.0202 | 0.0276 |
| 2. | **100KHz** | 0.0619 | 0.0276 | 0.0420 |
| 3. | **10KHz** | 0.1422 | 0.0552 | 0.0883 |
| 4. | **1KHz** | 0.3261 | 0.1288 | 0.2008 |
| 5. | **100Hz** | 0.8023 | 0.2952 | 0.4393 |
| 6. | **10Hz** | 2.340 | 0.6424 | 0.9475 |

TABLE III   Loss Factor at Specific Frequencies

| Sr. No. | tanδ | | | |
|---|---|---|---|---|
| | *Frequency* | *PSf* | *PSf + MWCNTs – Without Field* | *PSf + MWCNTs – With Field* |
| 1. | **1000KHz** | 0.0125 | 0.0076 | 0.0086 |
| 2. | **100KHz** | 0.0234 | 0.0103 | 0.0130 |
| 3. | **10KHz** | 0.0512 | 0.0203 | 0.0267 |
| 4. | **1KHz** | 0.1059 | 0.0455 | 0.0577 |
| 5. | **100Hz** | 0.2139 | 0.0952 | 0.1126 |
| 6. | **10Hz** | 0.4390 | 0.1736 | 0.1978 |

## IV. SUMMARY

Membranes are porous polymeric films and are routinely used for water treatment. The water permeability of these membranes can be improved by incorporating carbon nanotubes (CNTs) into the films and aligning them in direction of flow of water. It seems, it is possible to align CNTs in above films, if one adds magnetic nanoparticles in polymers solutions and casts the membrane in presence of magnetic field. It is seen that alignment of CNTs enhances the water permeability of membrane. The objective of the present paper was to find out if the dielectric properties of above membranes change on alignment of CNTs. The dielectric studies have been carried on Multiwall Carbon Nanotubes (MWCNTs) based Polysulfone (PSf) membranes. The dope solution used for synthesizing membranes consisted of PSf as base polymer, PVP as pore former, MWCNTs and $Fe_3O_4$ as additives and NMP as solvent. The membranes were casted in presence or absence of magnetic using indigenously developed membrane casting machine.

In all, three membranes, namely, Pure PSf membrane, (PSf / MWCNTs - Without Field) membrane and (PSf + MWCNTs - With Field) membrane were casted. The water permeation studies showed that nanotubes are aligned in (PSf / MWCNTs -With Field) membrane, which was casted in presence of magnetic field. The dielectric parameters (ε', ε'', and δ) for above three membranes were measured over a frequency range of 10Hz to 1MHz using and impedance analyzer. The results have been presented in form of plots and Tables. These studies show that real part ε' of dielectric constant for (PSf / MWCNTs - With Field) membrane is about 20% higher than that for (PSf / MWCNTs - Without Field) membrane. That is, dielectric constant of PSf / MWCNTs nanocomposite membrane increases when there was alignment of nanotubes. However, alignment of nanotubes did not have much effect on imaginary part ε'' of dielectric constant and the loss factor tan δ.

These studies suggest that dielectric measurements can be used for monitoring the degree of alignment CNTs in Polymer - CNTs composites. It is expected that higher is the degree of alignment, higher will be the value of ε'. The fact that that degree of alignment of nanotubes increases with the increase magnetic field used in casting the membrane, it is of interest to establish a correlation between the magnetic field H and dielectric constant ε'. Further, one would have to determine the degree of alignment independently using TEM or SAXS. In short, dielectric methods have potential of determining degree of alignment of nanotubes in a polymer matrix.

ACKNOWLEDGMENT

We thank Divya Padmanabhan, Priam V. Pillai, Sandeep M. Joshi and Avinash Vaidya for useful discussions and encouragement.

REFERENCES

[1] X. Y. Huang, B. Sun, Y. K. Zhu, S. T. Li, P. K. Jiang, "High-K Polymer Nanocomposites with 1D Filler for Dielectric and Energy Storage Application", Prog. Mater. Sci., vol. 100, 2019, pp. 187−225.

[2] X. Huang, P. Jiang, "Core-Shell Structured High-K Polymer Nanocomposites for Energy Storage and Dielectric Applications", Adv. Mater., vol. 27(3), 2015, pp. 546−554.

[3] K. Maex, M. R. Baklanov, S. H. Brongersma, Z. S. Yanovitskaya, "Low Dielectric Constant Materials for Microelectronics", *J. Appl. Phys.*, vol. 93, 2003, pp. 8793-8841.

[4] K. Yang, X. Huang, Y. Huang, L. Xie and P. Jiang, "Fluoro-Polymer@BaTiO$_3$ Hybrid Nanoparticles Prepared via RAFT Polymerization: Toward Ferroelectric Polymer Nanocomposites with High Dielectric Constant and Low Dielectric Loss for Energy Storage Application", Chem Mater., vol. 25(11), 2013, pp. 2327-2338.

[5] O. Kanoun, A. Bouhamed, R. Ramalingame, J. R. B. Quijano, D. Rajendran and A. Al-Hamry, "Review on Conductive. Polymer, CNTs Nanocomposite Based Flexible and Stretchable Strain and Pressure Sensors", Sensors, vol. 21(2), 2021, pp. 341-370.

[6] S. Yellampalli (ed.), Carbon Nanotube-Polymer Composites, InTech, Croatia, 2011, pp. 65-90.

[7] A. Sharma, B. Tripathi, Y.K. Vijay, "Dramatic Improvement in Properties of Magnetically Aligned CNT/ Polymer Nanocomposites", J. Memb. Sci., vol. 361, 2010, pp. 89–95.

[8] L. Nayak, M. Rahaman, D. Khastgir, T. K. Chaki, "Thermal and Electrical Properties of Carbon Nanotubes Based Polysulfone Nanocomposites", Polym. Bull., vol. 67, 2011, pp.1029–1044.

[9] R. W. Baker, Membrane Technology and Applications, 2nd ed., John Wiley & Sons, West Sussex, 2004, pp. 3-13.

[10] D. Das, A. Datta & A. Q. Contractor, "Various Types of Separation Membranes", Curr Sci, vol. 110, 2016, pp. 1426-1438.

[11] D. S. Sholl and J. K. Johnson (May, 2006). "Making high flux membranes with carbon nanotubes", Science 312, 2006, pp. 1003.

[12] R. Das, Carbon nanotubes for Clean Water, 1st ed., Springer International Publishing, Nature: Switzerland, 2018, pp. 127-150.

[13] Bhakti Hirani, P. S. Goyal, S. Kar and R. C. Bindal, "Polysulfone - Carbon nanotube composite ultrafiltration membranes for water purification" in Current Perspectives in Sustainable Environment Management, (Ed) S. Mishra, et al. (2017) pp. 76-81.

[14] B. Hirani, S. Kar, V. K. Aswal, R. C. Bindal and P. S. Goyal, "Polysulfone – CNT Composite Membrane with Enhanced Water Permeability", *AIP Conf. Proc.*, vol. 1942, 2018, pp. 050035.

[15] Bhakti Hirani and P.S. Goyal, Unpublished.

[16] P. V. D. Witte, J. Feijen, P. J. Dijkstra & J. W. A. V. D. Berg, "Phase Separation Processes in Polymer Solutions in Relation to Membrane Formation", J. Membr. Sci., vol. 117, 1996, pp. 1-31.

[17] P. S. Goyal and Bhakti Hirani, "Machine for Casting Thin Polymer Films Having Aligned Hollow Carbon Nanotubes Embedded in Them", Patent -Application No. 202021003922, dated 29-01-2020.

[18] B. Panda and P. S. Goyal, "Oleic Acid Coated Magnetic Nano-Particles: Synthesis and Characterizations", AIP Conf. Proc, vol. 1665, 2015, pp. 50020.